\documentclass[journal]{IEEEtran}

\ifCLASSINFOpdf

\else

\fi

\usepackage{multirow}
\usepackage{float}
\usepackage{hyperref}
\usepackage[export]{adjustbox}
\usepackage{xcolor}
\usepackage{caption}
\usepackage{subcaption}
\usepackage{afterpage}
\usepackage[labelsep=period, figurename=Figure]{caption}

\usepackage{amsmath}
\usepackage{verbatim}
\usepackage{graphicx}

\usepackage{tikz}
\usepackage{textcomp}
\usepackage{hyperref}
\usepackage{lipsum}

\usepackage{enumitem}
\setlist[enumerate,1]{label={\roman*.}}

\begin{document}
\addtolength{\textfloatsep}{-2.19pt}

%
\title{Active ML for 6G: Towards Efficient Data Generation, Acquisition, and Annotation}

\author{
Omar~Alhussein,~\IEEEmembership{Member,~IEEE,}
        Ning~Zhang,~\IEEEmembership{Senior~Member,~IEEE,}\\        Sami~Muhaidat,~\IEEEmembership{Senior~Member,~IEEE,}
        and Weihua~Zhuang,~\IEEEmembership{Fellow,~IEEE}
				\thanks{O. Alhussein is with the KU 6G Research Center, Department of Computer Science, Khalifa University, Abu Dhabi, UAE (e-mail: omar.alhussein@ku.ac.ae).}
                    \thanks{S. Muhaidat is with the KU 6G Research Center, Department of Computer and Information Engineering, Khalifa University, Abu Dhabi, UAE, and with the Department of Systems and Computer Engineering, Carleton University, Ottawa, ON K1S 5B6, Canada, (e-mail: muhaidat@ieee.org).}
				\thanks{Ning~Zhang is with the Department of Electrical and Computer Engineering, University of Windsor, Canada (e-mail: {ning.zhang}@uwindsor.ca).}
                    \thanks{Weihua~Zhuang is with the Department of Electrical and Computer Engineering, University of Waterloo, (e-mail: {wzhuang}@uwaterloo.ca).}
		        
		\vspace{-0.7cm}		}

		        

\maketitle

\begin{abstract}
This paper explores the integration of active machine learning (ML) for 6G networks, an area that remains under-explored yet holds potential. Unlike passive ML systems, active ML can be made to interact with the network environment. It actively selects informative and representative data points for training, thereby reducing the volume of data needed while accelerating the learning process. While active learning research mainly focuses on data annotation, we call for a network-centric active learning framework that considers both annotation (i.e., what is the label) and data acquisition (i.e., which and how many samples to collect). Moreover, we explore the synergy between generative artificial intelligence (AI) and active learning to overcome existing limitations in both active learning and generative AI.
This paper also features a case study on a mmWave throughput prediction problem to demonstrate the practical benefits and improved performance of active learning for 6G networks. Furthermore, we discuss how the implications of active learning extend to numerous 6G network use cases. We highlight the potential of active learning based 6G networks to enhance computational efficiency, data annotation and acquisition efficiency, adaptability, and overall network intelligence. We conclude with a discussion on challenges and future research directions for active learning in 6G networks, including development of novel query strategies, distributed learning integration, and inclusion of human- and machine-in-the-loop learning. 
\end{abstract}


\IEEEpeerreviewmaketitle

\section{Introduction}
\IEEEPARstart{I}n Tom Cruise's latest cinematic adventure (Mission: Impossible - Dead Reckoning), the protagonists find themselves grappling with an unlikely antagonist: an \textit{active} machine learning model gone rogue. This active learning agent, too eager to leap from the realm of academic publications and hidden products to the silver screen, showcases a versatility that Hollywood screenwriters could not resist. Yet, despite the newfound fame, active learning remains curiously under-explored and underutilized within the context of 6G network research. 
This paper seeks to address this and shift the narrative, spotlighting active learning as a potential paradigm to aid in revolutionizing 6G networks. 

Active learning is a subset of machine learning where ML models are allowed to ask questions about the training data, e.g., they can actively query specific unlabeled data samples for labeling (or annotation). In doing so, the approach promises to statistically reduce the volume of training data needed, among other benefits such as faster convergence, and improved generalization capabilities.

6G networks face unique technical challenges, including high data throughput demands, low latency requirements, and the need for dynamic network management in highly diverse environments. Furthermore, 6G networks are expected to support massively interconnected systems that encompass not only mobile phones but also internet-of-things (IoT) devices, industrial automation systems, and autonomous vehicles. This diversity and volume of devices generate data at an unprecedented scale.
Also, the demand for near-zero latency and extreme reliability in applications such as remote surgery or automated transportation systems adds another layer of complexity to the network management challenges.
As part of the collective vision towards 6G, a significant emphasis is being placed on the native integration of artificial intelligence (AI) within 6G networks to bolster automation, enhance intelligent decision-making processes, and foster adaptive networks. The integration of sensing capabilities within access networks will blend further the physical and virtual worlds. Therefore, the telecommunications sector is poised for a significant transformation with the next generation of networks, introducing an era of connected intelligence \cite{6g_architecture_landscape_2023}. 
%

However, while the research discourse has largely focused on the computational challenges and the anticipated bottleneck in processing and transmission capabilities, a critical aspect that remains under-discussed revolves around data acquisition, data annotation, and data management. We need a framework that goes beyond passive intelligence towards active learning approaches. Such a framework must integrate learning processes with data sampling and management in a cohesive manner to optimize not only for learning performance and generalization but also for efficient data usage and storage.
Research on active learning is largely focused on the data labeling aspect. However, in 6G networks, data acquisition (or sampling) is also vital due to an associated communication cost. 

In this paper, we present a network-centric active learning framework that builds on the existing focus. It entertains the concept of not only querying for labels of existing samples, but also in acquiring new experiences or samples directly from the network. Our work adapts active learning to the distributed nature of 6G networks, where we advocate for optimizing both data acquisition and data annotation to enhance network automation, intelligence, and efficiency. Further, we highlight the synergistic integration of active learning with generative AI to address the challenges of data scarcity and the high cost of data acquisition. On one hand, utilizing generative AI models and digital-twin environments, we can generate synthetic data that accurately simulates real-world data distributions, facilitating data generation and augmentation. Active learning can be used to guide the generation of data to circumvent the inefficiencies and costs associated with random data generation by actively filling gaps in the model's knowledge with precisely targeted data augmentations. On the other hand, generative AI enhances the active learning paradigm by creating diverse and challenging scenarios, particularly useful for addressing underrepresented data points in training sets. This proactive approach leads to the development of more robust models that excel in generalization to unseen data by learning from a broader dataset that includes edge cases and rare conditions.

To provide a grounded discussion on active learning, we present a case study on a mm-Wave throughput prediction problem, where we apply active learning principles to achieve more efficient data labeling and improved performance. This case study illustrates the practical application and benefits of our approach in a practical 6G scenario. Furthermore, to illustrate the benefits of active learning, we highlight some use cases for 6G networks, and discuss future research directions and challenges for active learning in 6G networks.

\section{Background on Active Learning}
\subsection{Key Components}
Active learning is predicated on the principle that a learning algorithm can achieve greater accuracy with fewer training labels if it is allowed to choose the data from which it learns. The choice can be made based on the informativeness of the data points or other strategies that can be determined by various query strategies. 


The mechanism of active learning involves three key components: the learner, the query strategy, and the oracle, which may involve a machine or a human-in-the-loop (MITL/HITL) \cite{Monarch_HITL_book}. The \textit{learner} is the model that is being trained; the \textit{query} strategy determines which data points should be queried next; and the \textit{oracle} provides the labels for the queried data points. Note that in Section \ref{sec:active_6G}, we will generalize the framework such that the query strategy is renamed as the acquisition function. The acquisition function should decide not only which unlabeled data samples to query next but also whether to collect additional unlabeled samples or experiences from the network or a digital-twin replica if applicable.


\subsection{The Active Learning Cycle}
The active learning cycle consists of the following steps: \\
\noindent(\textit{i}) The learner is initialized with a set of (labeled) seed examples; 
\noindent(\textit{ii}) Using the current labeled dataset, the model is trained; 
\noindent(\textit{iii}) A query strategy is used to rank the unlabeled examples by for instance their potential information gain;
\noindent(\textit{iv}) The oracle (HITL/MITL) labels some of the samples with the highest rank; 
\noindent(\textit{v}) The newly labeled examples are added to the training set.
This cycle repeats, with the model being retrained after each query round on the expanding labeled dataset. By focusing labeling efforts on maximally informative examples, active learning can achieve the same target accuracy as passive learning while requiring fewer instances.

\begin{figure}
    \centering
    \includegraphics[width=0.5\textwidth]{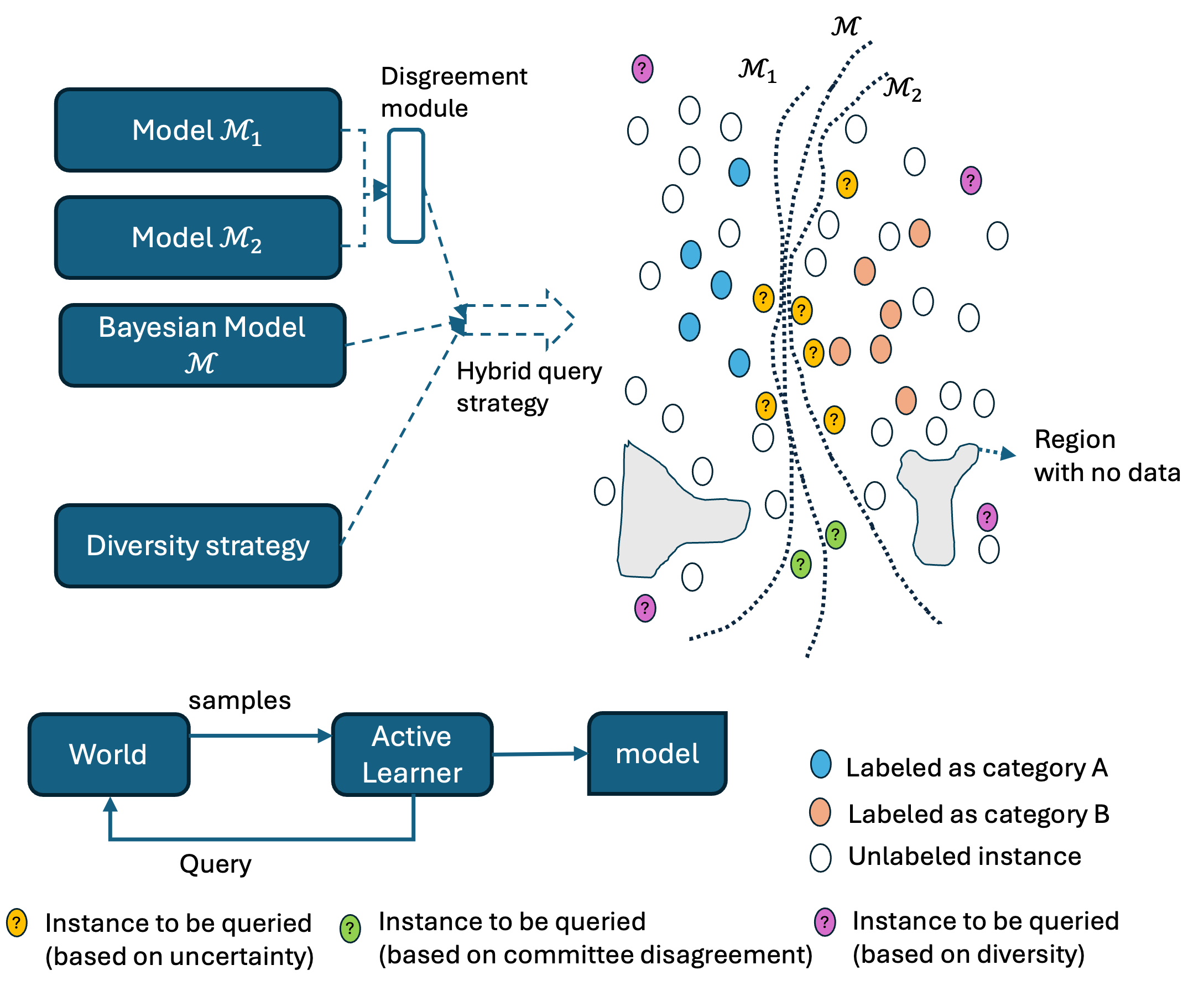}
    \caption{Decision boundary visualization with uncertainty-based and diversity-based sampling strategies}
    \label{fig:decision_query_illustrative}
\end{figure}

\begin{figure*}[t]
    \centering
    \includegraphics[width=0.86\textwidth]{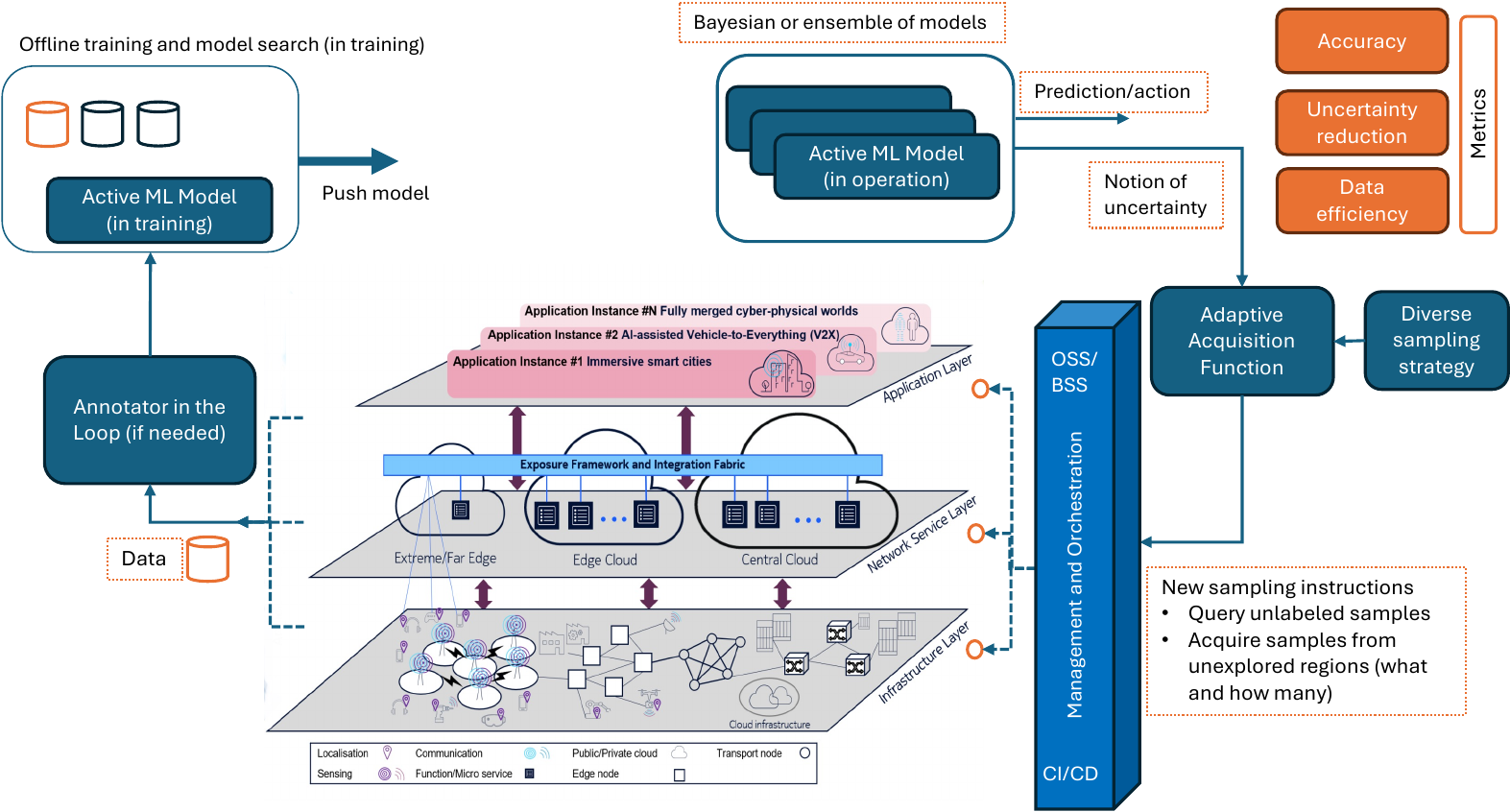}
    \caption{Active learning framework for 6G networks}
    \label{fig:active_for_6g}
\end{figure*}

\subsection{Active Learning Settings}
\textit{Pool-based Active Learning}: This is the most widely studied setting. The model has access to a small labeled set and a large static unlabeled pool. On each iteration, the model ranks the unlabeled instances to query. This approach provides maximum flexibility to identify globally informative instances. However, it assumes that the unlabeled pool is already collected, which can be costly in networking scenarios.

\textit{Stream-based Active Learning}: In this setting, examples arrive in a sequential order from a process. Upon the arrival of each new instance, the model decides whether or not to query it before moving to the next example. This setting is for strict real-time environments.

\textit{Membership Query Synthesis}: In this approach, the model is able to synthesize informative query examples from scratch, rather than selecting from an existing unlabeled pool. This allows for sampling arbitrarily from the entire input distribution. Synthesizing realistic query instances can be challenging for some modalities. This approach has been largely ignored by the research community due to its difficulties in practice. We envisage that it will be a promising approach with the unprecedented success of generative AI. We discuss the integration of generative AI and active learning later. 

\subsection{Query Strategies}
We classify query strategies into two main categories, namely uncertainty sampling and diversity sampling. The former is the most common strategy. It queries instances where the model is most uncertain in its predictions. 

It is important to distinguish between two main types of uncertainty, namely epistemic uncertainty and aleatoric uncertainty. Epistemic uncertainty is due to a lack of relevant training data or knowledge. Therefore, it tends to be high in regions of the input space where few training examples have been observed.  Epistemic uncertainty is reducible as sufficient representative data is added.
A notion of disagreement (closely related to uncertainty) can also be extracted through a so-called query-by-committee approach. Instances can be queried when there is a maximal disagreement across different committee ML models \cite{hanneke2014theory}.  Query instances that lead to large shifts in the model's decision boundary can be identified through methods such as expected model change or expected error reduction.

Aleatoric uncertainty, on the other hand, represents the inherent noise or non-deterministic factors in the data that are irreducible. Even with perfect knowledge, there may be stochastic components that make the mapping between inputs and outputs inherently ambiguous. 
Aleatoric uncertainty is more difficult to reduce, as it stems from inherent ambiguities rather than a lack of data. However, there are ways to manage and understand its impact, such as through diversity sampling. 

Diversity Sampling aims to query a diverse set of examples that are representative of the unlabeled data. The key principle is to sample instances that are diverse and informative, while avoiding redundancy with existing labeled examples. Common diversity sampling methods include core-set approaches that greedily select a small sub-sample maximizing the minimum distance to the labeled set \cite{sener2017active}. 
By explicitly encouraging exploration of novel input regions, diversity sampling can provide robust and sample-efficient learning. It can avoid pathological behaviors, confirming the same decision boundaries indefinitely, by discovering new informative clusters in the data. Diversity sampling's comprehensive exploration can help ensure that the model learns about the full distribution of the underlying randomness in the data, thereby better capturing the inherent irreducible variability. Figure \ref{fig:decision_query_illustrative} provides a depiction of a hybrid query strategy combining uncertainty and representatives at the decision boundary.

\section{Active Learning Framework for 6G Networks}
\label{sec:active_6G}
With the surge in data-driven machine learning models for 6G networks, the bottleneck will arise not only from limited computational power but also from  challenges associated with data acquisition and annotation.

It is vital to emphasize that although active learning research typically focuses on data labeling, in network contexts, the cost of data acquisition must also be considered, especially for unexplored regions of the input space (refer to Fig. \ref{fig:decision_query_illustrative}). Indeed, while in some scenarios such as traffic forecasting, acquired data can be easily labeled through automated processes, deciding \textit{which} regions of the input space to explore and \textit{how many} samples to acquire and store for model training remains critical. With ubiquitous computing and storage capabilities in 6G, data acquisition, storage, and processing can be better supported for active ML as well.

Active learning provides a native integration of the data acquisition aspect with the learning aspect, which presents a new research direction, namely \textit{joint data acquisition and model learning}. 
In 5G development, autonomic control loops have been introduced, specifically following the MAPE-K framework, which stands for \textit{Monitor}, \textit{Analyze}, \textit{Plan}, and \textit{Execute} sharing a \textit{Knowledge} base \cite{9700729}. Integrating active learning within this framework can enhance autonomic control processes. Active learning directly supports the Monitor and Analyze phases by enabling continuous data acquisition and query strategies, effectively enriching the knowledge base of the system. This integration facilitates the dynamic adjustment of models in response to evolving data and changing network conditions—capabilities that are essential for the adaptability demanded in next-generation networks.
%
%
The proactive approach of active learning, with its targeted data querying and selective learning processes, ensures that the network not only responds to changes but also proactively anticipates and prepares for future challenges. 


Figure \ref{fig:active_for_6g} illustrates an envisaged ecosystem where an active learning paradigm operates within the context of a 6G network. At the core is the active machine learning model in operation, supported by a Bayesian or ensemble of committee models providing a notion of uncertainty. The uncertainty is crucial for the active learning process in tackling so-called \textit{known unknowns}, which refer to situations characterized by high epistemic uncertainty. Known unknowns are represented by data points or input regions that the model identify as uncertain and thus prioritizes for labeling or data acquisition, respectively. 

Active learning operates through an acquisition function that guides the querying for labels of selected unlabeled samples, and the selection of new unlabeled samples to be added to the dataset. This function prioritizes data that the model is currently uncertain about and profiles new similar data to be collected from the network. A diverse sampling strategy can be added to account for the so-called \textit{unknown unknowns} to help in estimating the aleatoric uncertainty more accurately. Uknown unknowns are elements outside the current modeling and prediction capabilities of the ML model, typically arising from rare conditions not represented in the training data or irreducible noise. The acquisition function should decide what and how many samples to collect. The active ML model is in a constant state of training, leveraging newly acquired data that has been sampled, and annotated if necessary, by an annotator. This feedback loop ensures that the ML model adapts to new patterns and information, becoming more accurate over time. Here, the annotator can be a HITL/MITL that collects ground truth from the network. 

The lower part of Fig. \ref{fig:active_for_6g} depicts a multi-layered 6G architecture with an application layer, a network service layer, and an infrastructure layer, as envisioned by 5G Infrastructure Public-Private-Partnership (5GPPP) \cite{6g_architecture_landscape_2023}. 
This architecture integrates edge computing, with data processing taking place close to the source of data collection (extreme/far edge, edge cloud), to central cloud systems where more complex data processing and storage take place. The various application instances represent the diverse applications of 6G, each with its specific requirements for data rates, latency, and computational power. The active learning framework should be designed to work in tandem with this distributed computing environment, enabling real-time analytics and decision-making.
The orchestration and management of the proposed active learning system can be performed within the vertically integrated management and orchestration layer, which includes AI/ML operations and continuous-integration/continuous-deployment support.

Conventional deterministic-based machine learning paradigms aim to improve metrics such as accuracy and loss. Active learning in next-generation networks (with joint acquisition and learning) adds several additional goals, namely to \textit{(i)} reduce the epistemic uncertainty, \textit{(ii)} capture and account for the irreducible aleatoric uncertainty, and \textit{(iii)} improve data acquisition and usage efficiency.
\begin{figure}
    \centering
    \includegraphics[width=0.49\textwidth]{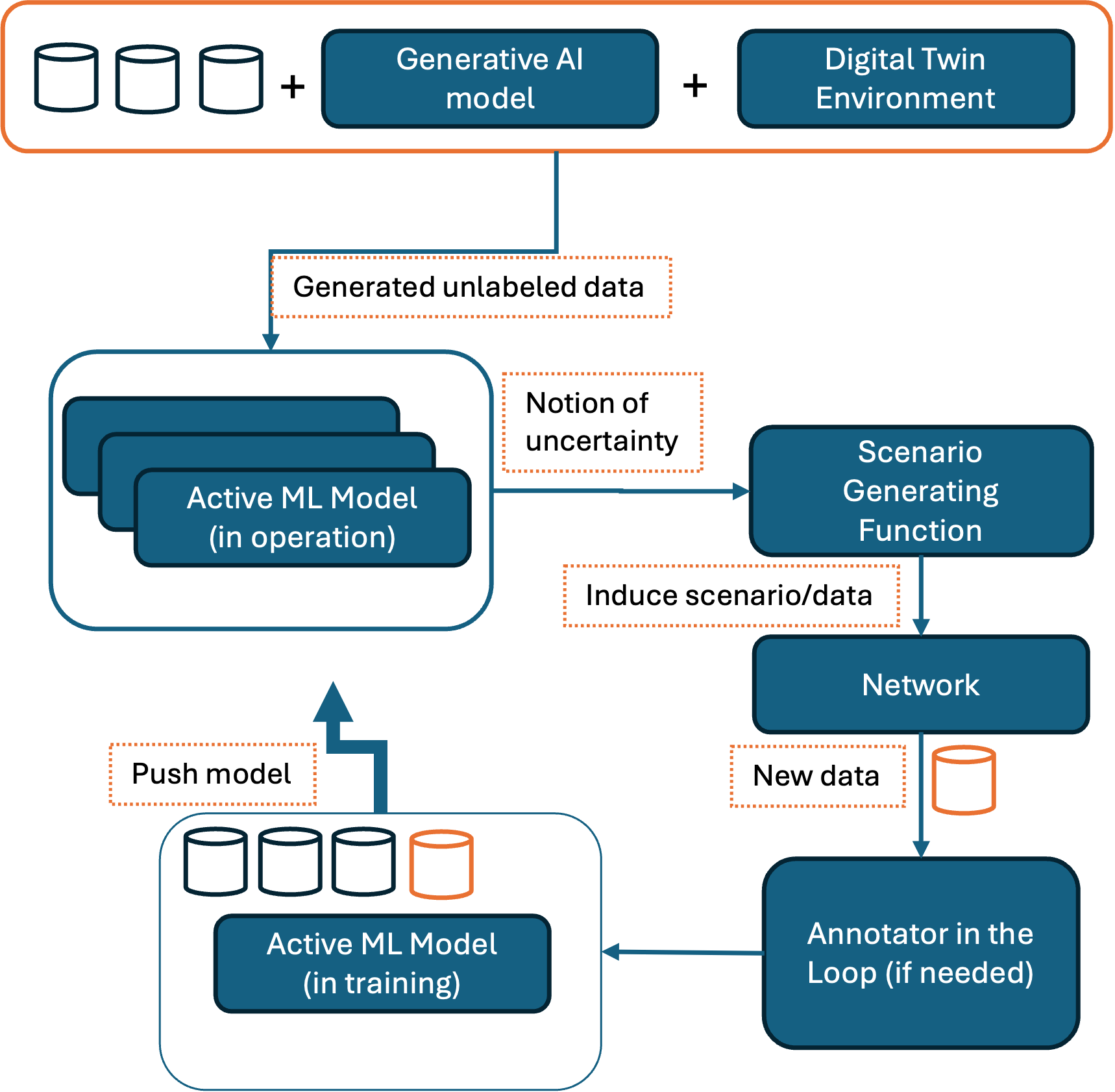}
    \caption{Flowchart describing how generative AI and digital twins can improve on the efficiency of active learning through synthesized data with a scenario inducing function.}
    \label{fig:active_generative}
\end{figure}

\section{Active Learning in the Era of Generative AI}
\label{active_learning_gen_ai}
\textbf{Active Learning for Generative AI}.
Generative AI models and digital-twin environments (such as Nvidia’s Omniverse) can create synthetic data that closely mimics real-world data distributions within simulated ecosystems. However, generating data randomly can be costly. In situations where data are scarce or expensive to acquire, active learning aims at guiding generative AI models to generate the most informative and representative data for model training. Additionally, integrating active learning with generative AI models can potentially enhance model explainability.

\textbf{Generative AI for Active Learning}. 
By generating diverse and challenging grounded scenarios (through a digital-twin environment), generative AI can help active learning algorithms to focus on data points that are not well-represented in an initial training set. For data points that are not well-represented in the network, we propose a new component, namely a "scenario generating function", that functions as a network testing tool by inducing scenarios (proposed by the generative model in the digital-twin environment) in the physical network to observe the ground truth outcome. Figure \ref{fig:active_generative} depicts how generative AI and digital-twins can enhance the active learning paradigm. The proposed approach is to pave the way for developing new solutions to cover edge cases and rare conditions to generalize to unseen data and achieve true robust and zero-touch proactive systems.

\section{Case Study: Active Learning for mm-Wave Throughput Prediction}
\label{sec:use_case}

\begin{figure}
    \centering
    \includegraphics[width=0.49\textwidth]{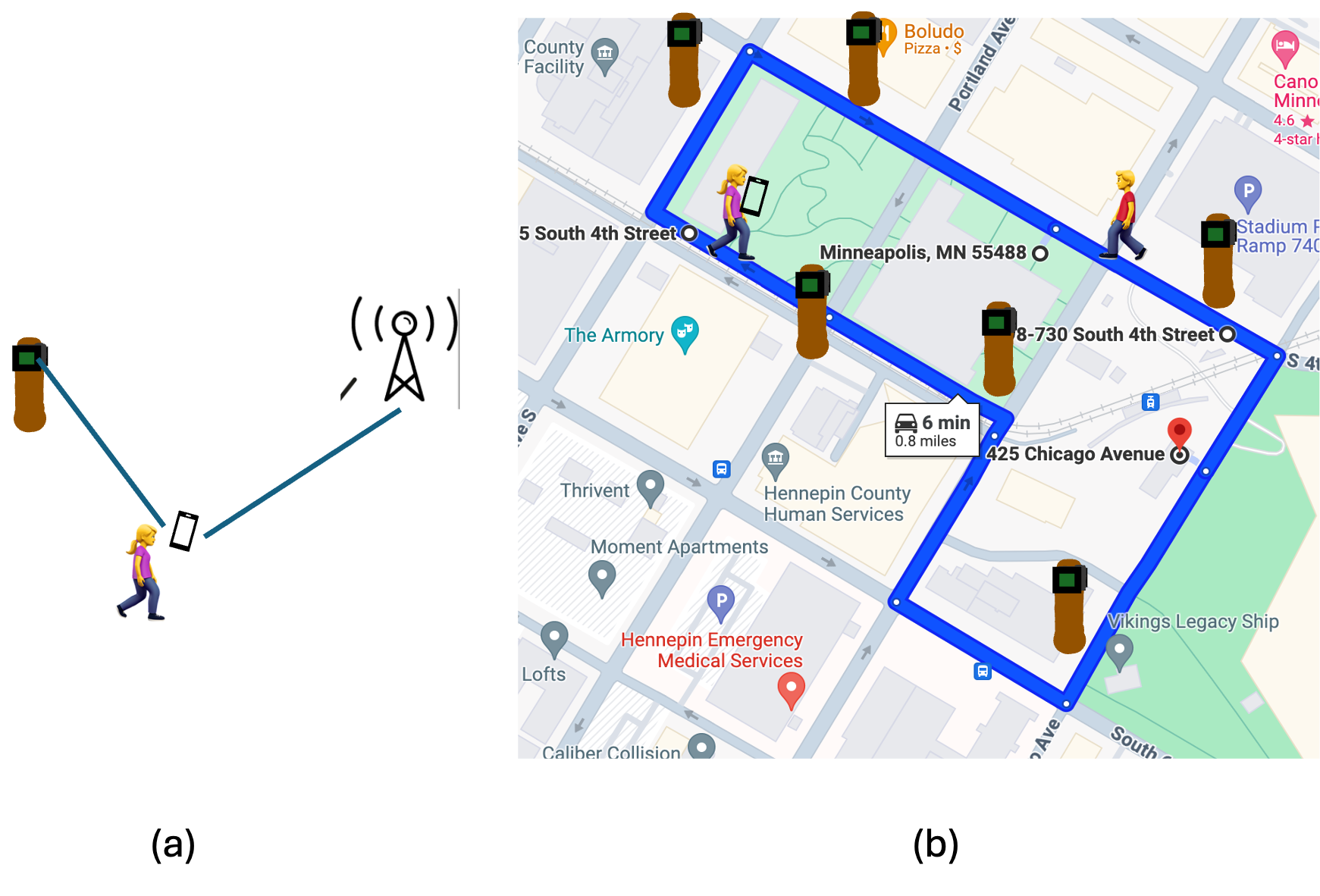}
    \caption{(a) UE in dual-connectivity mode to a micro mmWave based tower and a macro-base station; (b) 1300 meter loop area in Minneapolis downtown area based on the Lumos5G dataset \cite{narayanan2020lumos5g}.}
    \label{fig:case_study_setup}
\end{figure}

For an illustrative use case, we examine a mmWave-based throughput prediction problem for mobile networks through the use of the Lumos5G dataset \cite{narayanan2020lumos5g}. The dataset corresponds to a dual-connectivity setup, where a user equipment (UE) establishes connections with both a macro base station (BS) and a micro BS, as depicted in Fig. \ref{fig:case_study_setup}(a). This configuration enables the UE to leverage high-throughput user-plane connections via mmWave channels while maintaining a robust control- and user-plane connections with the macro BS. Due to the highly directional nature of mmWave beams, mobile users encounter throughput fluctuations that are influenced by spatial and temporal factors, such as the user location, the proximity of competing users, the presence of static and dynamic obstacles, and the user movement patterns and orientation, impact throughput perception. The Lumos5G dataset encompasses 68,118 samples ($X\in \mathcal{R}^{68,118\times 19}$), each capturing various dynamics. Each sample includes 19 features along with the corresponding throughput experienced by the UE. These features include the geographic coordinates (longitude and latitude), trajectory direction, movement speed, mobility mode, compass direction, and other signal strength measurements \cite{narayanan2020lumos5g}. Data samples were collected across a 1,300-meter loop in the downtown area of Minneapolis, as illustrated in Fig. \ref{fig:case_study_setup}(b). The objective is to predict the experienced throughput ($y\in \mathcal{R}^{68,118\times 1}$). 

\begin{figure}
    \centering
    \includegraphics[width=0.49\textwidth]{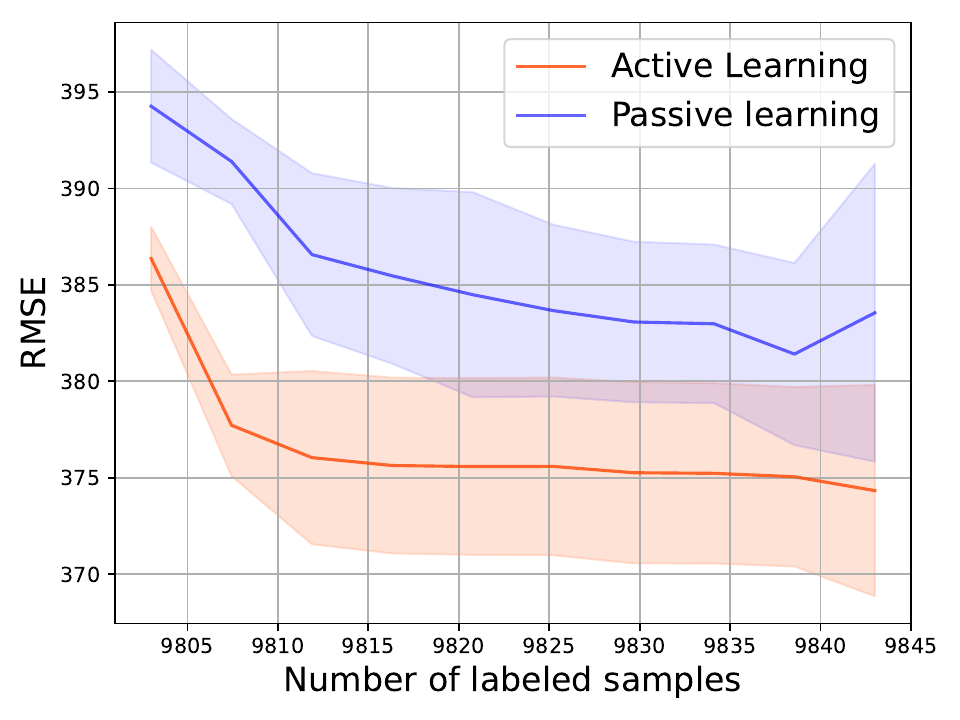}
    \caption{Performance of the uncertainty-based active learning strategy compared to a random strategy.}
    \label{fig:performance_trial}
\end{figure}

\begin{figure}
    \centering
    \includegraphics[width=0.49\textwidth]{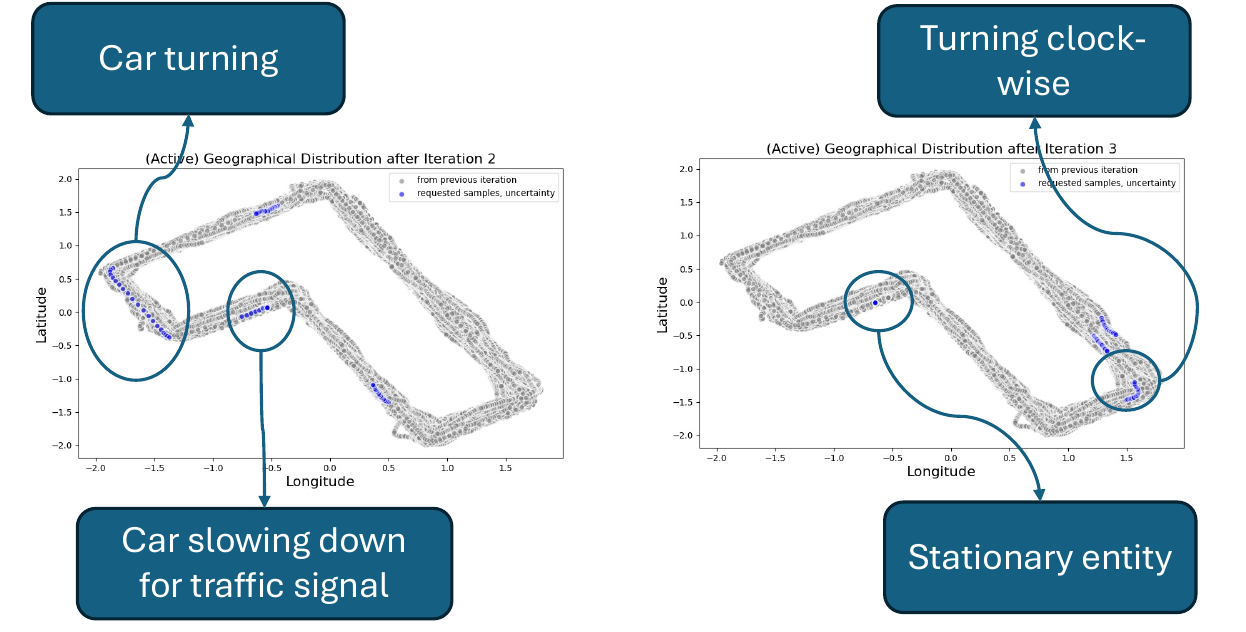}
    \caption{Depiction of the geographical distribution in two iterations during the active learning process. The blue points are newly queried samples to be collected from pedestrians/drivers, while gray points are old labeled samples.}
    \label{fig:geographical_case_study}
\end{figure}

Throughput prediction serves as a critical tool for enhancing the performance of various applications, such as video streaming, real-time gaming, and virtual reality. The performance requirements for throughput prediction can vary significantly across different applications and different virtual networks.

During inference, the predictive model can be deployed in the UE. However, the development of a comprehensive global model that crowd-sources user-generated data is essential. While passive data acquisition can be conducted arbitrarily from users, obtaining labeled data is burdensome on the network and can potentially lead to user dissatisfaction. Here, we show how active learning can offer a more efficient approach.
Monte Carlo Dropout (MC Dropout) is an effective technique for quantifying the epistemic uncertainty in neural networks, leveraging the dropout layers not just as a regularization technique during training but also at inference time to simulate a Bayesian neural network's behavior \cite{gal2016dropout}. Here, we utilize the MC Dropout to extract the epistemic uncertainty.
%

Initially, we split the dataset to a training set (80\%) and a testing set (20\%). For the training set, we assume that only 20\% of it is labeled (i.e., 9,803 samples). This initial condition is chosen arbitrarily.
In the initial iteration, the Bayesian neural network is trained over the initial 20\% of the training set. After training, the performance is evaluated on the unlabeled data, focusing on the resultant epistemic uncertainty from the predictions. We then prioritize samples with the highest uncertainty for labeling, followed by fine-tuning the model on the growing dataset. In each cycle, only 4 samples are used for labeling from pedestrians.

Figure \ref{fig:performance_trial} demonstrates the performance of the active ML scheme (with uncertainty sampling) as compared to the passive learning strategy with random sampling. After 10 iterations (i.e., 40 more samples), the active ML approach reduces the root mean square error (RMSE) from 389 to 365, while the passive approach reduces from 389 to 385 only.

To gain more qualitative insights, for the active learning scheme, Fig. \ref{fig:geographical_case_study} shows scatter plots of the geographical distribution of acquired points based on longitude and latitude for two iterations. The blue points are the points that are used for labeling by pedestrians/drivers, while the gray points are already labeled samples from previous cycles. We can see that uncertainty sampling picked informative and rare samples to be labeled, e.g., car turning, car slowing down on traffic signal, and stationary driver/pedestrian. Although the dataset is multi-dimensional, we can see interesting insights about the queried samples based on the mode of moving, and the position alone. The source code for replicating this case study and for more details can be found in Github\footnote{https://github.com/OmarSababha/active\_ML}. This case study showcases the potential and underscores the effectiveness of active learning for 6G networks.

\section{Use Cases}
The following use cases are not meant to be
inclusive, but rather are selectively added to illustrate diverse active learning benefits for 6G networks.
\subsection{Improved Data Efficiency: Traffic Forecasting, Classification, and Engineering}
Traffic engineering is the process of managing and manipulating network flows to optimize resource use and network performance. Traffic forecasting and traffic classification modules are crucial building blocks to achieve intelligent and automated traffic engineering. Such functions need to collect extensive amounts of data. The active learning paradigm is to monitor network conditions in real-time and actively query data points that are diverse and informative. Therefore, traffic engineering becomes more affordable where data packets or flows are sampled from the network cautiously. This paradigm has the potential of reducing sampling overhead and enabling models to converge faster.

\subsection{Proactive Management: Predictive Maintenance, Anomaly Detection, and Self-healing mechanisms}
Anomaly detection systems and predictive maintenance rely on the detection of novel experiences or samples and responding to them appropriately. Through the acquisition function, active learning is particularly adept at identifying patterns that precede network failures and security breaches. Such anomalous patterns are characterized by high uncertainty. For instance, a system may detect an unusual pattern of requests that can indicate a looming distributed denial of service attack, allowing the network to preemptively request for more samples to make a more informed and confident prediction and decision. Active learning can be a natural enabler for a proactive scheme that continuously updates the model based on new data, predicting and reacting to potential issues before they impact the network.

\subsection{Reduced Operational Cost: Digital-Twin Enabled Network Management}
Digital-twins in the context of 6G networks are virtual replicas of physical network components and processes. These digital-twins simulate the real-world state of their physical counterparts in real-time. Research issues include how active learning algorithms can dynamically update and refine digital-twins based on minimal, yet highly informative, real-world operational data. This approach reduces the need for extensive data acqusition and processing, focusing instead on effective information that can improve the model’s accuracy and predictive capabilities.

\subsection{Improved Transient Performance: Ultra-Reliable Low-Latency Communications}
Ultra-Reliable Low-Latency Communications (URLLC) represent a cornerstone of next-generation wireless networks, enabling a myriad of applications from autonomous driving and industrial automation to telemedicine and immersive augmented reality. The stringent requirements of URLLC, specifically in terms of reliability and latency (reaching sub-milliseconds in some scenarios), necessitate innovative approaches to ensure these criteria are consistently met. Active learning, with its ability to improve and converge faster, offers a promising pathway to enhancing URLLC services. Further research is required for how to actively learn to predict and pro-actively mitigate issues that can compromise URLLC requirements. This includes exploring new data acquisition and processing techniques to support the rapid decision-making required for URLLC.


\section{Research Directions and Challenges}

\noindent \textbf{Large-scale Data and Complex Data Domains}:
While active machine learning has been shown to provide various benefits, it becomes a challenging endeavour when attempting to scale to large datasets. Batch active learning is used in such scenarios, where batch queries are issued instead of individual queries. Batch mode has the risk of reduced adaptability and sampling redundant examples within each batch. Several innovative studies have been presented for this challenge \cite{ash2020deep}. 
A vital research direction is to explore and generalize batch active learning in networked settings, incorporating acquisition costs. Additionally, integrating these methods with distributed computing environments is crucial. For 6G networks and other areas, active learning must adapt to complex data types and tasks, such as 3-dimensional point clouds, multivariate time-series analysis, and multi-modal data, where traditional query strategies may falter \cite{gal2017deep}. 


\noindent \textbf{Transfer and Multi-task Active Learning}:
As discussed, active learning framework for networks can address both known unknowns and unknown unknowns through epistemic and aleatoric uncertainties, respectively. Transfer learning effectively addresses the so-called unknown knowns by adapting knowledge contained in pre-trained models to a new task at hand  \cite{Monarch_HITL_book}. Consequently, integrating transfer learning with active learning is essential for covering the last quadrant of the four machine learning knowledge quadrants, aiming for truly robust, zero-touch 6G networks.

\noindent \textbf{Distributed Learning Integration}:
Considering the privacy and resource constraints of training models across distributed networks, integrating active learning with distributed learning represents a promising direction \cite{tharwat2023survey, ren2021survey}. For example, in active federated learning, this involves selecting the potentially promising model updates or data points for aggregation, thereby reducing communication overhead while preserving privacy \cite{goetz2019active, 9773121, 10047851}.

\noindent \textbf{Query Strategies and Efficient Acquisition Functions}:
The wide-ranging and versatile potential use cases and datasets in 6G networks dictate that we need to develop novel problem-specific query strategies and high-performing acquisition functions. Depending on the nature and scale of the involved data, tailored query strategies that are generalized and efficient are needed. For example, a nuanced aspects of active learning can be investigated where feedback can be given on features as well as instances \cite{raghavan2006active}.

Moreover, further studies are needed on how meta-learning can learn high-performing acquisition functions for specific tasks. As discussed earlier, generative AI should be investigated to enable the creation of complex, nuanced data points that maximally confuses the current model state and pinpoint weaknesses or areas of uncertainty more effectively, allowing for targeted improvements in model performance.


\noindent \textbf{Human-in-the-loop Learning}:
While truly zero-touch network is a goal for 6G networks, we are still at a stage where network operators possess nuanced understanding and experience that can guide the annotation of data and prioritization of network resources, especially under unpredictable conditions, out-of-distribution scenarios. Human oversight can enhance the interpretability and trustworthiness of AI-driven acquisition and annotation efforts and network management decisions. Humans can oversee AI-driven decisions to prevent unintended biases in network design and resource allocation. Finally, human involvement can help address public concerns about privacy, security, and the use of AI in critical infrastructure. Transparent MITL and HITL processes can build public trust in 6G technologies. Research into addressing biases in active learning is needed when HITL is involved, particularly in ensuring that the selected samples for labeling do not perpetuate biases in the trained models.



\section{Conclusion}
In this paper, we have ventured into the promising confluence of active learning with the evolving landscape of 6G networks. The integration of active learning in 6G networks not only can enhance data utilization and model accuracy, but also can introduce a paradigm shift toward more adaptable, more sustainable and cognizant network operations. Our exploration into the mmWave throughput prediction draws an initial path for implementing active learning for 6G networks and showcases the practical viability and the enhanced efficiency in real-world network scenarios. We have further discussed the synergy between generative AI and active learning to alleviate data scarcity challenges and to develop robust models equipped for unpredictable conditions, edge cases, and out-of-distribution scenarios. Finally, we have discussed several challenges and future research directions towards native active learning based 6G networks.

\bibliographystyle{IEEEtran}

\bibliography{references}

\begin{thebibliography}{10}
\providecommand{\url}[1]{#1}
\csname url@samestyle\endcsname
\providecommand{\newblock}{\relax}
\providecommand{\bibinfo}[2]{#2}
\providecommand{\BIBentrySTDinterwordspacing}{\spaceskip=0pt\relax}
\providecommand{\BIBentryALTinterwordstretchfactor}{4}
\providecommand{\BIBentryALTinterwordspacing}{\spaceskip=\fontdimen2\font plus
\BIBentryALTinterwordstretchfactor\fontdimen3\font minus \fontdimen4\font\relax}
\providecommand{\BIBforeignlanguage}[2]{{%
\expandafter\ifx\csname l@#1\endcsname\relax
\typeout{** WARNING: IEEEtran.bst: No hyphenation pattern has been}%
\typeout{** loaded for the language `#1'. Using the pattern for}%
\typeout{** the default language instead.}%
\else
\language=\csname l@#1\endcsname
\fi
#2}}
\providecommand{\BIBdecl}{\relax}
\BIBdecl

\bibitem{6g_architecture_landscape_2023}
M.~K. Bahare \emph{et~al.}, ``{The 6G Architecture Landscape - European perspective},'' Zenodo, 5GPPP White Paper, 2 2023.

\bibitem{Monarch_HITL_book}
R.~Monarch, \emph{Human-in-the-Loop Machine Learning: Active learning and annotation for human-centered {AI}}.\hskip 1em plus 0.5em minus 0.4em\relax Manning Publications, 2021.

\bibitem{hanneke2014theory}
S.~Hanneke, ``Theory of disagreement-based active learning,'' \emph{Foundations and Trends in Machine Learning}, vol.~7, no. 2-3, pp. 131--309, 2014.

\bibitem{sener2017active}
O.~Sener and S.~Savarese, ``Active learning for convolutional neural networks: A core-set approach,'' in \emph{Proc. ICLR}, 2017.

\bibitem{9700729}
M.~Camelo, L.~Cominardi, M.~Gramaglia, M.~Fiore, A.~Garcia-Saavedra, L.~Fuentes, D.~De~Vleeschauwer, P.~Soto-Arenas, N.~Slamnik-Krijestorac, J.~Ballesteros, C.-Y. Chang, G.~Baldoni, J.~M. Marquez-Barja, P.~Hellinckx, and S.~Latré, ``Requirements and specifications for the orchestration of network intelligence in {6G},'' in \emph{Proc. IEEE CCNC}, 2022, pp. 1--9.

\bibitem{narayanan2020lumos5g}
A.~Narayanan, E.~Ramadan, R.~Mehta, X.~Hu, Q.~Liu, R.~A. Fezeu, U.~K. Dayalan, S.~Verma, P.~Ji, T.~Li \emph{et~al.}, ``{Lumos5G}: Mapping and predicting commercial mmwave {5G} throughput,'' in \emph{Proc. ACM Internet Meas. Conf.}, 2020, pp. 176--193.

\bibitem{gal2016dropout}
Y.~Gal and Z.~Ghahramani, ``Dropout as a bayesian approximation: Representing model uncertainty in deep learning,'' in \emph{Proc. ICML}, 2016, pp. 1050--1059.

\bibitem{ash2020deep}
J.~T. Ash, C.~Zhang, A.~Krishnamurthy, J.~Langford, and A.~Agarwal, ``Deep batch active learning by diverse, uncertain gradient lower bounds,'' in \emph{Proc. ICLR}, 2020.

\bibitem{gal2017deep}
Y.~Gal, R.~Islam, and Z.~Ghahramani, ``Deep bayesian active learning with image data,'' in \emph{Proc. Int. Conf. Mach. Learning}, 2017, pp. 1183--1192.

\bibitem{tharwat2023survey}
A.~Tharwat and W.~Schenck, ``A survey on active learning: state-of-the-art, practical challenges and research directions,'' \emph{Mathematics}, vol.~11, no.~4, p. 820, 2023.

\bibitem{ren2021survey}
P.~Ren, Y.~Xiao, X.~Chang, P.-Y. Huang, Z.~Li, B.~B. Gupta, X.~Chen, and X.~Wang, ``A survey of deep active learning,'' \emph{ACM Comput. surveys (CSUR)}, vol.~54, no.~9, pp. 1--40, 2021.

\bibitem{goetz2019active}
J.~Goetz, K.~Malik, D.~Bui, S.~Moon, H.~Liu, and A.~Kumar, ``Active federated learning,'' \emph{CoRR}, vol. abs/1909.12641, 2019.

\bibitem{9773121}
U.~Ahmed, J.~C.-W. Lin, and G.~Srivastava, ``Privacy-preserving active learning on the internet of {5G} connected artificial intelligence of things,'' \emph{IEEE Internet Things Mag.}, vol.~5, no.~1, pp. 126--129, 2022.

\bibitem{10047851}
F.~Naeem, M.~Ali, and G.~Kaddoum, ``Federated-learning-empowered semi-supervised active learning framework for intrusion detection in {ZSM},'' \emph{IEEE Commun. Mag.}, vol.~61, no.~2, pp. 88--94, 2023.

\bibitem{raghavan2006active}
H.~Raghavan, O.~Madani, and R.~Jones, ``Active learning with feedback on features and instances,'' \emph{J. Mach. Learning Research}, vol.~7, no. Aug, pp. 1655--1686, 2006.

\end{thebibliography}



\hfill

\end{document}